\documentclass[sigconf]{acmart}
\usepackage[T1]{fontenc}
\usepackage{xcolor}
\usepackage{graphicx}
\usepackage{listings}
\usepackage{tikz-uml}
\usepackage{comment}
\usepackage{xspace}
\usepackage{url}
\usepackage{tabularx}
\usepackage{fancyvrb}

\definecolor{commentColor}{RGB}{100, 100, 100}
\definecolor{stringColor}{RGB}{150, 0, 0}
\newcommand{\Symbol}[0]{\texttt{Symbol}\xspace}
\newcommand{\Qutes}[0]{\textsf{Qutes}\xspace}

\lstdefinelanguage{Qutes}{
  morekeywords={int, bool, string, qubit, quint, qustring, float, void, return, *, /, \%, +, -, not, and, or, by, swap, pauliy, payliz, grover, mcz, mcx, mcy, mcp, hadamard, measure, print, println, barrier, !, =, <, >, var, for, foreach, search, in, where, if, else, while, do},
  morecomment=[s]{/*}{*/},
  morecomment=[l]//,
  morestring=[b]",
  morestring=[b]'
}

\lstdefinestyle{qutes} {%
  basicstyle={\footnotesize\ttfamily},   
  frame=tb,
  captionpos=b
  xleftmargin={0.75cm},
  numbers=left,
  stepnumber=1,
  firstnumber=1,
  numberfirstline=true,	
  identifierstyle=\color{olive},
  keywordstyle=\color{blue}\bfseries,
  ndkeywordstyle=\color{editorGreen}\bfseries,
  stringstyle=\color{stringColor}\ttfamily,
  commentstyle=\color{commentColor}\ttfamily,
  language=Qutes,
  alsodigit={.:;},	
  tabsize=2,
  showtabs=false,
  showspaces=false,
  showstringspaces=false,
  extendedchars=true,
  breaklines=true,
}

\begin{document}

\title{Qutes: A High-Level Quantum Programming Language for Simplified Quantum Computing}

\author{Simone Faro}
\email{faro@unict.it}
\affiliation{%
  \institution{Università degli studi di Catania}
  \city{Catania}
  \country{Italia}
}
\author{Francesco Pio Marino}
\email{francesco.marino@phd.unict.it}
\affiliation{%
  \institution{Università degli studi di Catania}
  \city{Catania}
  \country{Italia}
}
\affiliation{%
  \institution{Univ Rouen Normandie, INSA Rouen Normandie, Université Le Havre Normandie, Normandie Univ, LITIS UR 4108}
  \city{Rouen}
  \country{France}
}
\author{Gabriele Messina}
\email{gabriele.messina@studium.unict.it}
\affiliation{%
  \institution{Università degli studi di Catania}
  \city{Catania}
  \country{Italia}
}

\setcounter{secnumdepth}{5}

\begin{abstract}
Quantum computing leverages the principles of quantum mechanics to perform computations far beyond the capabilities of classical systems, particularly in fields such as cryptography and optimization. However, current quantum programming languages often require low-level implementation, posing significant barriers for many developers due to their steep learning curve and limited abstraction. In response, we introduce \textbf{Qutes}, a high-level quantum programming language designed to simplify quantum algorithm development while maintaining the flexibility required for advanced applications. By abstracting complex quantum operations and allowing intuitive expressions through high-level constructs, Qutes enables users to write efficient quantum programs without extensive knowledge of quantum mechanics or circuit design. Built upon Qiskit, Qutes translates its syntax directly into executable quantum code, facilitating seamless integration with quantum hardware. This paper provides an overview of the language's architecture, core functionalities, and its ability to unify classical and quantum operations within a single framework. Additionally, we demonstrate Qutes' application in key quantum algorithms, showcasing its potential to make quantum programming more accessible and practical for a wider range of developers and researchers.
\end{abstract}
\maketitle

\section{Introduction}

Quantum computing represents a transformative leap in computational science, leveraging the principles of quantum mechanics to solve problems that are intractable for classical computers. By utilizing quantum bits (qubits), which harness phenomena such as superposition and entanglement, quantum computers can process information in fundamentally novel ways. This paradigm shift opens doors to groundbreaking advancements across fields such as cryptography, artificial intelligence, optimization, and materials science.

As quantum computing progresses, the need for sophisticated quantum programming languages becomes increasingly critical. These languages act as a bridge between high-level algorithm design and the low-level requirements of quantum hardware, enabling the effective exploitation of quantum devices. While languages like Qiskit, Cirq, and others have significantly advanced quantum programming, they often reveal limitations in abstraction, usability, and integration as the field evolves. These challenges highlight the necessity for innovative solutions that address the growing complexity of quantum systems while maintaining accessibility for developers.

The rapid proliferation of quantum computing tools and resources reflects the expanding interest from both academia and industry. Over fifty open-source projects now offer diverse libraries, frameworks, and simulators that support quantum algorithm development and experimentation. These initiatives, such as those cataloged in \cite{Software}, span a range of applications, from theoretical algorithm design to hardware-specific optimizations. Simulators, in particular, play a pivotal role by allowing developers to prototype and refine quantum algorithms on classical systems before deployment on actual quantum hardware~\cite{Simulators}.

Despite this progress, existing tools face significant challenges. Many quantum programming languages remain low-level~\cite{HSKRH24}, requiring expertise in quantum mechanics, circuits, and gates. This steep learning curve often alienates developers from non-specialized backgrounds, limiting the accessibility of quantum programming. Consequently, there is a growing demand for high-level languages that abstract the underlying complexities, making quantum development more intuitive and inclusive.

To address these gaps, we introduce \textbf{Qutes}, a high-level quantum programming language designed to simplify quantum development while retaining the flexibility and power necessary for advanced applications. Qutes abstracts the intricacies of quantum gates and circuits, enabling developers to focus on high-level algorithmic constructs. Unlike traditional languages that demand detailed knowledge of quantum mechanics, Qutes provides an approachable interface for quantum computation, fostering innovation across a wider audience.

Qutes also enhances quantum-classical integration, a key requirement for hybrid workflows in fields like machine learning. Built on the solid foundation of Qiskit, Qutes translates high-level syntax into executable quantum code, facilitating seamless execution on real quantum hardware. This approach not only simplifies quantum programming but also bridges the gap between classical and quantum computing environments, a critical step toward practical quantum applications.

This paper explores the design and implementation of Qutes, presenting its architecture, core functionalities, and example applications, including Grover’s search and the Deutsch-Jozsa algorithm. By demonstrating the utility of Qutes, we aim to showcase its potential to make quantum programming more accessible, bridging the divide between high-level algorithm design and hardware implementation. The source code for \textbf{Qutes} is freely available on GitHub\footnote{Current version available at \url{https://github.com/GabrieleMessina/qutes_lang}}, enabling easy adoption across cloud-based platforms such as Colab or GitHub Codespaces.

The remainder of this paper is structured as follows: Section~\ref{sec:sota} provides a comprehensive review of the state of the art in quantum programming languages, with a focus on the most desirable features for quantum software development. Section~\ref{sec:qutes:arch} explores the architecture of Qutes in detail, highlighting its key components and design principles. In Section~\ref{sec:qutes_lang}, we present illustrative examples of well-known quantum algorithms implemented using Qutes, showcasing its practicality and versatility. Finally, Section~\ref{sec:conc} summarizes our findings and offers concluding remarks.

\section{State of the Art in Quantum Programming Languages}\label{sec:sota}

This section surveys the current landscape of quantum programming languages, focusing on their capabilities, limitations, and the specific challenges they aim to address. By examining established frameworks, alongside innovative languages. We highlight the evolution of quantum programming tools and their role in shaping the future of quantum computing.

\subsubsection*{Qiskit}
Qiskit~\cite{qiskit} is one of the most widely adopted quantum programming frameworks, developed by IBM. It provides a comprehensive Python-based toolkit for creating, simulating, and executing quantum circuits. The modular structure of Qiskit enables developers to build quantum circuits, run simulations, and interact with IBM's quantum hardware through a unified interface. Qiskit excels in offering a low-level interface to quantum operations, giving users granular control over circuit design and optimization. Advanced features, such as custom gates, noise modeling, and integration with machine learning libraries, make Qiskit a powerful tool for both research and real-world applications. However, its steep learning curve and intricate syntax can be challenging for developers unfamiliar with quantum mechanics. Despite these challenges, Qiskit’s extensive documentation and active community support mitigate its accessibility issues, making it a go-to choice for professionals and researchers alike.

\subsubsection*{Cirq}
Cirq~\cite{cirq} is a quantum programming framework developed by Google, designed to interface seamlessly with Google's quantum processors. It provides users with the tools to construct quantum circuits programmatically and execute them on both simulators and real quantum hardware. Cirq is particularly well-suited for researchers and developers working on quantum error correction and hardware-specific optimizations, as it allows fine-grained control over circuit execution. A notable feature of Cirq is its support for customizing gate decompositions and scheduling, enabling efficient utilization of hardware resources. Although its low-level syntax provides flexibility, it can be less approachable for developers seeking high-level abstractions. Nevertheless, Cirq’s strong integration with Google's Sycamore processors and its focus on hardware-specific features make it a valuable tool for advancing quantum computing research.

\subsubsection*{Quipper}
Quipper~\cite{quipper} is a functional programming language for quantum computing that emphasizes scalability and modularity. It is particularly suited for implementing large-scale quantum algorithms, such as Shor’s factoring algorithm and Grover’s search. Unlike procedural languages, Quipper adopts a functional programming paradigm, offering developers powerful compositional tools for building complex circuits. Quipper also supports higher-order functions and type safety, ensuring robust and error-free code. Its ability to efficiently compile and execute large quantum circuits has made it a preferred choice for algorithmic research. However, its reliance on functional programming concepts poses a steep learning curve for developers accustomed to procedural or object-oriented paradigms. Despite this limitation, Quipper remains a benchmark language for theoretical and experimental quantum algorithm design.

\subsubsection*{Silq}
Silq~\cite{silq} is a high-level quantum programming language that introduces groundbreaking features such as automatic uncomputation to simplify quantum programming. This feature allows Silq to automatically revert temporary quantum states, eliminating the need for developers to manage ancilla qubits manually. Silq also boasts an intuitive syntax, making it accessible to programmers without deep quantum expertise. By abstracting away many low-level details, Silq enables the development of quantum algorithms with greater speed and accuracy. Despite its promising capabilities, Silq is still in its early stages of development and lacks the broad hardware integration offered by more established frameworks such as Qiskit and Cirq. As a result, its use is currently limited to theoretical exploration and simulation rather than practical deployment on quantum hardware.

\subsubsection*{Twist}
Twist~\cite{twist} is another innovative high-level quantum programming language, specifically designed to tackle challenges related to entanglement management and variable purity in quantum programs. Unlike other languages, Twist enforces a rigorous treatment of entanglement, ensuring that unintended interactions between qubits are avoided. This approach is particularly useful for applications requiring precise quantum state control, such as quantum cryptography and secure communications. Twist also provides type safety mechanisms that help prevent common programming errors, such as inadvertent quantum state overwriting. While its specialized features make Twist a powerful tool for advanced quantum programming, they may not appeal to developers looking for general-purpose solutions or high-level abstractions.

\subsubsection*{OpenQASM}
Open Quantum Assembly Language (OpenQASM)~\cite{openqasm} is a hardware-agnostic assembly language designed to provide a low-level representation of quantum circuits. Its simple, assembly-like syntax allows developers to describe quantum operations and execute them on various quantum hardware platforms. OpenQASM has been instrumental in standardizing quantum circuit representation, enabling compatibility across different hardware backends. The third iteration, OpenQASM 3, expands its capabilities to include quantum-classical hybrid operations, making it suitable for a wider range of applications. By offering a direct and portable interface to quantum hardware, OpenQASM remains a cornerstone in the quantum programming ecosystem, widely used in both academia and industry.

\subsubsection*{Microsoft Q\#}
Q\#~\cite{qsharp} is Microsoft’s quantum programming language that integrates seamlessly with classical programming workflows. As part of the Microsoft Quantum Development Kit, Q\# provides extensive libraries, debugging tools, and integrated quantum simulators. Developers can design hybrid quantum-classical algorithms using familiar classical programming constructs, such as loops and conditionals, while leveraging Q\#’s robust quantum features. Q\# also supports seamless integration with Microsoft’s Azure Quantum platform, enabling cloud-based execution of quantum programs on real hardware. Its strong type system and intuitive syntax make it a popular choice for both educational and professional use. However, its adoption is primarily centered around the Microsoft ecosystem, which may limit its appeal for users seeking broader platform compatibility.

\subsubsection*{PyQuil}
PyQuil~\cite{pyquil} is a Python-based library developed by Rigetti Computing for creating and running quantum programs. PyQuil utilizes the Quil language to define quantum circuits and offers extensive support for hybrid quantum-classical programming. Its architecture allows seamless integration with Rigetti’s Quantum Cloud Services (QCS) and local quantum virtual machines, making it highly flexible. PyQuil includes tools for noisy quantum simulations and parameterized circuits, enabling developers to experiment with near-term quantum algorithms on NISQ (Noisy Intermediate-Scale Quantum) devices. Additionally, PyQuil’s integration with Python makes it an accessible and versatile tool for developers familiar with Python-based data science workflows. Its focus on hybrid workflows positions it as a strong candidate for exploring practical quantum applications.

\subsection{Desirable Features in Quantum Computing Programming Languages}

The development of a quantum computing programming language requires a well-thought-out set of features to ensure usability, scalability, and efficiency. This section discusses the key attributes that such a language should encompass, aiming for a general perspective applicable across different implementations.

\subsubsection*{Quantum/Classic Collaboration}
Quantum computing languages must allow seamless interaction between quantum and classical computing paradigms. Such collaboration is essential for enabling operations like quantum measurements, conditional statements, and loops that interact with classical data. While quantum systems often rely on probabilistic outcomes, efficient data exchange between the classical and quantum realms is critical for practical applications. This requires innovative mechanisms to facilitate the integration of classical and quantum logic without overcomplicating the development process.

\subsubsection*{Purity Management and Type Systems}
Purity management in quantum programming refers to the distinction between pure and mixed quantum states. This feature is essential for ensuring that operations adhere to the principles of quantum mechanics while optimizing resource utilization. Advanced type systems that explicitly define quantum states and operations improve both code clarity and error prevention. For example, type annotations can differentiate between quantum states and classical variables, ensuring that operations are valid within the constraints of quantum mechanics.

\subsubsection*{Natural Language and Fluent Syntax}
The syntax of a quantum programming language should strike a balance between simplicity and functionality. Adopting intuitive syntax resembling natural language or leveraging a fluent programming style can make the language accessible to a wider audience. Such design considerations reduce the learning curve, especially for developers without a strong background in quantum computing. However, care must be taken to avoid overloading the language with constructs that sacrifice readability or introduce unnecessary complexity.

\subsubsection*{Data Types and Structures}
A diverse set of data types is crucial for quantum programming. This includes classical types such as integers, floating-point numbers, and strings, alongside quantum-specific types such as qubits and superpositions. Additionally, robust data structures, including arrays and dictionaries, allow developers to organize and manipulate quantum and classical data effectively. A well-designed language should support extensibility, enabling the introduction of user-defined types and structures as needed.

\subsubsection*{Arithmetic and Quantum Operations}
Providing basic arithmetic and logical operations on quantum data types is a fundamental requirement for a quantum computing language. Such operations should abstract low-level circuit and gate implementations, allowing developers to focus on algorithmic design rather than hardware details. In addition to arithmetic, the language should include quantum-specific operations, such as unitary transformations and measurements, in an intuitive and user-friendly manner.

\subsubsection*{Library Support and Abstraction}
Comprehensive library support is a cornerstone of an effective programming language. Libraries that implement common quantum algorithms, such as Grover’s and Shor’s algorithms, significantly enhance productivity. Abstractions that hide underlying complexities, such as qubit topology or gate arrangements, allow developers to work at a higher level of abstraction, making quantum computing more accessible to non-experts.

\subsubsection*{Prior Knowledge Requirements}
To widen the reach of quantum computing, programming languages should minimize the need for prior knowledge of quantum mechanics or specific programming paradigms. A user-friendly language should abstract complex concepts without compromising functionality, enabling individuals from diverse backgrounds to engage with quantum programming. This approach democratizes access to quantum computing and fosters innovation across disciplines.

\subsubsection*{Automatic Uncomputation}
Uncomputation is a vital feature in quantum programming, ensuring that intermediate quantum states are safely discarded without introducing errors. Automating this process is crucial to simplifying the development of complex algorithms. A well-designed language should provide built-in mechanisms to handle uncomputation seamlessly, reducing the cognitive load on developers while adhering to quantum mechanical principles.

\subsubsection*{Linearity and Non-Cloning Compliance}
Linearity, a fundamental property of quantum mechanics, ensures that operations do not violate the no-cloning theorem. Quantum programming languages must incorporate constructs that enforce linearity, such as borrow-use patterns or linear type systems. These features ensure that quantum resources are managed correctly, preventing inadvertent duplication of quantum states and preserving the integrity of computations.

\subsubsection*{Hardware Compatibility}
The ability to run code on existing quantum hardware is a critical feature of a quantum programming language. Compatibility with a variety of quantum processors, as well as simulators for development and testing, ensures that the language remains practical and versatile. Hardware-independent abstractions further enhance usability, allowing developers to write portable code without being tied to specific hardware implementations.

\subsubsection*{Scalability and Future-Readiness}
A forward-looking quantum programming language should be designed with scalability and adaptability in mind. As quantum technology evolves, the language must support new algorithms, data structures, and operational paradigms. Extensibility and modularity are key attributes that enable a language to remain relevant and effective in a rapidly changing technological landscape.

Each of these features represents a critical aspect of a quantum programming language's design, collectively forming the foundation for a tool that is both powerful and accessible. By incorporating these principles, a programming language can meet the diverse needs of quantum computing developers and researchers.

\subsection{Comparative Analysis of Quantum Programming Languages}

This section synthesizes the collective strengths and limitations of some well known quantum programming languages, focusing on their unique implementations of key attributes.

One of the most critical features, \textit{quantum-classical collaboration}, sees diverse levels of support across languages. High-level tools like Q\# and PyQuil excel in integrating quantum and classical paradigms, enabling seamless workflows for hybrid algorithms. By contrast, low-level languages like OpenQASM and Quipper focus on direct control of hardware or circuits, limiting their utility for hybrid applications. Silq and Twist introduce innovative mechanisms for this collaboration, but with caveats—Silq’s hybrid features remain theoretical, while Twist restricts its quantum-classical interaction to specific entanglement and purity use cases. Qutes strikes a middle ground by supporting type promotion between classical and quantum types and enabling quantum-to-classical conversions via measurement.

\textit{Purity management}, another cornerstone of quantum computation, is explicitly supported by Silq and Twist, which enforce linearity and manage entanglement to preserve quantum state integrity. Other languages, including Qiskit and Cirq, lack built-in features for purity handling, relying on developer oversight to ensure compliance with quantum mechanical principles. This gap illustrates the trade-off between advanced theoretical guarantees and practical usability. Qutes prioritizes practical usability, with no built-in mechanisms for purity management. Instead, it provides robust operations like automatic measurement to simplify the conversion of quantum states into classical values when needed.

Regarding \textit{syntax usability}, Silq stands out for its natural language-inspired structure, lowering the barrier to entry for newcomers. Twist and Q\# also aim for user-friendliness but retain some complexity, especially for advanced entanglement or type management. In contrast, Qiskit, Cirq, and OpenQASM emphasize low-level operations, which provide granular control but introduce significant learning curves for developers unfamiliar with quantum mechanics. Qutes combines an accessible syntax with a structured approach to integrating quantum and classical programming constructs, making it approachable for new developers while offering sufficient depth for complex applications.

\textit{Data type and structure support} also varies significantly. Qiskit and Q\# offer a rich array of quantum-specific and classical data types, supporting modularity and scalability. PyQuil and Cirq provide moderate support, focusing on types most relevant to near-term quantum devices. In contrast, Quipper and OpenQASM remain highly restrictive, limiting types to those directly tied to quantum hardware or theoretical constructs. Qutes introduces versatile data types, including qubit, quint, and qustring for quantum operations, and supports arrays of both classical and quantum data types, emphasizing flexibility and compatibility for a wide range of use cases.

In terms of \textit{automatic uncomputation}, Silq and Twist lead the field with automated mechanisms to manage ancilla qubits and discard temporary states. This feature reduces cognitive load and errors in algorithm development. Other languages, such as Q\# and PyQuil, implement uncomputation manually, while OpenQASM and Quipper largely overlook this aspect. Qutes facilitates practical management by embedding measurement-based conversion into its control structures but leaves uncomputation tasks to the developer, aligning with its balanced approach to usability and flexibility.

\textit{Hardware compatibility} is a key strength of Qiskit, PyQuil, and Q\#, which support a wide range of quantum processors, making them practical for real-world applications. Cirq aligns closely with Google’s Sycamore hardware, offering fine-grained optimizations. Twist and Silq are currently limited to simulations, emphasizing theoretical advancements over direct hardware integration. OpenQASM’s hardware-agnostic approach ensures broad compatibility but sacrifices abstraction and usability. Qutes leverages Qiskit to maintain hardware-agnostic capabilities, with a focus on extensibility and support for a wide range of simulators and quantum devices as part of its development roadmap.

Finally, \textit{library support} distinguishes high-level tools from their lower-level counterparts. Qiskit, Q\#, and PyQuil provide extensive libraries for common algorithms and error mitigation techniques, accelerating development. In contrast, languages like Quipper and OpenQASM lack such abstractions, focusing instead on circuit-level programming. Qutes, while nascent, includes common quantum operations as built-in language features rather than relying on external libraries. It is designed to encourage modularity and the development of higher-level constructs in future updates.

In summary, high-level languages such as Q\#, PyQuil, and Silq prioritize abstraction and developer accessibility, making them suitable for hybrid and practical applications. Meanwhile, lower-level options like OpenQASM and Quipper cater to hardware control and theoretical research. Qutes bridges these domains, offering a balance of abstraction, quantum-classical integration, and hardware flexibility, making it an adaptable option for a variety of quantum programming tasks. As the field progresses, the convergence of features such as automatic uncomputation, advanced type systems, and hardware compatibility will likely define the next generation of quantum programming languages.

\section{A Brief Overview of Qutes' Architecture and Design}\label{sec:qutes:arch}
\Qutes is a Domain-Specific Language (DSL) designed to offer developers a high-level interface for quantum programming by abstracting and simplifying the underlying complexities of quantum code. It achieves this by transpiling DSL instructions directly into Qiskit~\cite{qiskit} code, thereby leveraging Qiskit's robust features and enabling the execution of quantum programs on actual quantum hardware. This approach allows developers to benefit from Qiskit’s capabilities while focusing on high-level program design without needing to manage low-level quantum operations.

The \Qutes grammar was implemented in Python, utilizing the ANTLR~\cite{antlr} tool to define syntax and parse rules. Once a source file is parsed, the resulting Abstract Syntax Tree (AST) is traversed to instantiate symbols, represented by instances of a custom Python class, \Symbol. Each \Symbol object encapsulates essential information, including type and scope, required for the DSL to function effectively. 

After the initial symbol instantiation, a second pass through the syntax tree is conducted. During this iteration, quantum operations are translated into corresponding quantum circuit instructions, while non-quantum operations are executed directly in Python, ensuring efficient handling of classical operations within the DSL. To manage this process, the \texttt{QuantumCircuitHandler} class plays a pivotal role by logging all quantum operations specified by the user.

As the second traversal concludes, the \texttt{QuantumCircuitHandler}, aggregates these operations, generating a \texttt{QuantumCircuit} instance that incorporates all necessary \texttt{QuantumRegisters} associated with declared variables. This final step allows the handler to assemble the entire sequence of quantum operations, creating a cohesive circuit that mirrors the user's intent within the DSL. Through this structured approach, \Qutes effectively encapsulates both quantum and classical functionality, enabling a seamless transition from high-level instructions to a fully-realized quantum circuit.

When a quantum variable interacts with a classical one, \Qutes automatically initiates a measurement operation. This measurement records the quantum state, saving the result in the associated classical variable, thereby enabling interoperability between the quantum and classical contexts. Conversely, when a classical variable is assigned to a quantum variable, the \texttt{TypeCastingHandler} encodes the classical value directly into the quantum circuit, ensuring that the quantum system accurately reflects the specified classical state.

The \texttt{TypeCastingHandler} plays a central role in managing these interactions. It dynamically determines the appropriate operations for translating values between classical and quantum realms, ensuring type consistency and enabling seamless transitions between classical and quantum data representations. By handling these conversions, the \texttt{TypeCastingHandler} ensures that mixed classical-quantum operations are intuitive and efficient, maintaining coherence in assignments throughout the \Qutes DSL environment.

Control flow structures such as branches and loops are available within \Qutes and are designed to work with classical Boolean conditions as their parameters. When a quantum variable is supplied as a condition parameter, the associated quantum registers are automatically measured, and the result is used to evaluate the Boolean condition. This approach provides a straightforward way to incorporate quantum states within classical logic flows, though it requires that quantum data be cast into the classical form for compatibility with standard control structures.

To assist users in these cases, \Qutes offers type-casting functionality that allows quantum variables to be easily converted to classical Boolean values when required. This type-casting enables smooth integration of quantum data into decision-making logic, making it intuitive for developers to implement conditional constructs based on quantum states without manual measurement operations.

In addition to these control structures, \Qutes provides utilities such as arrays and functions, enhancing the DSL's versatility. Basic operations on both classical and quantum data types are supported, giving developers flexibility in manipulating data and implementing custom logic. This combination of control flow, data manipulation utilities, and type compatibility ensures that \Qutes remains a powerful, accessible tool for hybrid classical-quantum programming, bridging the gap between high-level program design and quantum circuit implementation.

\section{Type System in Qutes}

The type system in Qutes is designed to seamlessly integrate classical and quantum computing paradigms, enabling developers to work efficiently across both domains. This section outlines the core elements and functionalities of the type system, highlighting its role in simplifying quantum-classical hybrid programming.

Qutes distinguishes itself by supporting a comprehensive set of data types that cater to both classical and quantum programming needs. These types form the foundation of the language's operations and are essential for implementing complex quantum algorithms.

For classical data types, Qutes includes standard types commonly found in many programming languages. The \texttt{bool} type represents boolean values, allowing logical conditions to be expressed and evaluated. The \texttt{int} type supports integer values, useful for counters and discrete computations. Floating-point numbers are handled by the \texttt{float} type, enabling precise arithmetic operations. Text and sequences of characters are managed using the \texttt{string} type, ensuring versatility in handling classical data representations.

For quantum-specific data types, Qutes introduces specialized constructs that align with the requirements of quantum computing. The \texttt{qubit} type represents a single quantum bit, the fundamental unit of quantum information. Collections of qubits, such as quantum registers, are encapsulated in the \texttt{quint} type, which supports operations on multiple quantum states simultaneously. The \texttt{qustring} type allows for the representation of quantum strings, limited to bitstrings due to current hardware constraints, facilitating the management of quantum sequences.

The operations supported in Qutes are similarly categorized into quantum and classical domains. Quantum operations include fundamental gates such as the Hadamard and Pauli gates, alongside phase gates for manipulating quantum states. Higher-level quantum operations, such as superposition addition and cyclic permutation, simplify the implementation of advanced quantum algorithms. Classical operations, on the other hand, encompass arithmetic operations like addition, subtraction, multiplication, and division, as well as logical and comparison operators such as AND, OR, greater-than, and less-than. These operations ensure robust functionality for classical computations while maintaining compatibility with quantum systems.

Variables in Qutes are always passed by reference, ensuring efficient data handling in both classical and quantum contexts. The seamless cooperation between classical and quantum paradigms is a standout feature of Qutes. Classical variables can be promoted to quantum equivalents through type promotion, enabling their direct use in quantum circuits. Conversely, quantum variables can be converted to classical values through a measurement process, which collapses the quantum state into a definite classical value. This bidirectional compatibility allows developers to bridge the classical and quantum domains effortlessly.

Arrays in Qutes further enhance its versatility by supporting collections of elements that may be either classical or quantum types. These arrays allow indexed access, enabling developers to read or modify elements using their position within the array. The ability to iterate through arrays simplifies complex operations across multiple elements, making it easier to manage both classical and quantum data structures.

The support for functions in Qutes promotes modular and reusable code. Functions can accept multiple parameters and return values, accommodating both classical and quantum types as inputs and outputs. This capability facilitates the implementation of complex quantum algorithms, ensuring flexibility in program design.

Control structures such as conditional statements and loops are integral to the Qutes programming model. Branching logic is supported through \texttt{if} and \texttt{if-else} statements, while repeated execution is enabled by \texttt{while} loops. Iteration through arrays is streamlined by the \texttt{for-each} loop, simplifying repetitive tasks. Notably, control structures in Qutes require classical values for conditions. When quantum variables are used as conditions, they are automatically measured to produce a classical value, ensuring compatibility with standard programming constructs.

In conclusion, the type system in Qutes strikes a delicate balance between expressiveness and simplicity. By providing a robust framework for managing both classical and quantum data types, it empowers developers to explore the full potential of quantum-classical hybrid programming. The comprehensive support for diverse data types, operations, and control structures ensures that Qutes remains a powerful and accessible tool for the next generation of quantum software development.

\section{A Brief Showcase of Qutes' Capabilities}\label{sec:qutes_lang}
Although the project remains a work in progress, it already incorporates several advanced features, including support for quantum-specific types and arrays, user-defined functions, and implicit type casting between compatible quantum and classical types, which can induce measurements on the circuit as required.
%
%
The code example depicted in Figure \ref{fig_grammar_syntax} illustrates the definition of quantum variables and vectors containing quantum states, including superpositions of values. Additionally, the code demonstrates addition operations between quantum variables, which clearly implement circuits for quantum register addition. This example showcases the straightforward syntax of \Qutes for handling complex quantum data structures and operations. Figure \ref{fig_substring} shows an example demonstrating how the \Qutes language natively implements Grover's search algorithm through instructions that allow substring searching. In both examples shown, the evaluation of a quantum variable—whether for verifying its value or for printing—requires a measurement operation, which collapses its state into one of the possible values the variable represents. All of these implementation details are transparent to the programmer.

The following provides a selection of additional examples illustrating how the new language can be effectively used to implement more complex procedures, such as quantum entanglement propagation and the Deutsch-Jozsa algorithm.

\subsubsection*{Cyclic shift of a quantum register}
A cyclic shift is an operation that rotates the positions of elements within a register, moving each qubit to the right (or left) by a specified number of positions, $k$. Traditionally, a cyclic shift operation on a classical machine requires linear time relative to the size of the register, as each qubit must be sequentially repositioned. 
In \Qutes, however, the cyclic shift is implemented using a dedicated instruction (see Figure \ref{fig_grammar_syntax}) based on the rotation algorithm developed by Faro, Pavone, and Viola \cite{FPV24}. This algorithm enables a quantum register to be rotated in constant time, resulting in significant efficiency improvements for large quantum registers. The \Qutes cyclic shift instruction translates directly into a quantum circuit that completes the rotation in a constant number of steps, making it highly efficient compared to classical implementations.

\subsubsection*{Entanglement propagation}
The example presented in Figure \ref{fig_esp} uses the entanglement swap protocol \cite{Zangi_2023} to perform an entanglement propagation along an array of qubits. It is a fundamental quantum communication technique that allows entanglement to be established between two qubits that have never directly interacted. This is achieved by initially preparing two pairs of entangled qubits. A qubit from each pair is then brought together and measured using a Bell measurement. While this measurement destroys the local entanglement within the pairs, it effectively transfers the entanglement to the two remaining qubits, which have not been in direct contact.
The circuit structure for the entanglement swap protocol involves preparing the entangled pairs, performing a Bell measurement on two intermediary qubits, and applying corrections to the final qubits based on the measurement results. In the provided \Qutes example, this protocol is extended to propagate entanglement along an array of qubits, ultimately producing an entangled state between the first and last qubits of the array.

\subsubsection*{The Deutsch-Jozsa Algorithm}
The Deutsch-Jozsa algorithm \cite{Deutsch1992RapidSO} is a fundamental quantum algorithm that determines whether a function is constant or balanced. In \Qutes, this algorithm is implemented with remarkable simplicity (see Figure \ref{fig_dj}). The input function, which is guaranteed to be either constant or balanced, is defined explicitly to accept a quantum register. By initializing the input qubits in a superposition state and setting the output qubit to $|-\rangle$, the Deutsch-Jozsa algorithm operates as expected. At the end of the computation, the output registers are evaluated to reveal whether the function is balanced or constant. This streamlined syntax in \Qutes makes it easy to implement high-level quantum algorithms without delving into complex circuit details.

\section{Progressive Pathways and Future Directions}\label{sec:conc}

While the current implementation of \Qutes demonstrates notable capabilities in quantum programming, several avenues for further research and refinement have been identified. Expanding the language’s operational framework to encompass a wider suite of fundamental quantum operations, including arithmetic (e.g., addition, multiplication) and comparative functions is critical for broadening its practical applicability. In parallel, establishing methods to export \Qutes code to widely used quantum programming languages, particularly \textsf{Qiskit} and \textsf{QASM}, would markedly improve interoperability and facilitate integration within existing quantum workflows. Developing a comprehensive standard library containing essential quantum functions and algorithms is also indispensable for streamlining application development within the \Qutes ecosystem. Additionally, the creation of quantum specific debugging tools remains a priority to support precise and efficient testing tailored to quantum computational contexts. Further research into generalizing Grover’s algorithm for application in database operations governed by arbitrary filter functions, as well as introducing native operations for calculating the maximum and minimum of a set, would greatly enhance its versatility and extend its applicability to a wider range of complex data queries. Moreover, focusing on automatic uncomputation, incorporating linear variables, and ensuring variable purity as implemented in Twist~\cite{twist} are paramount to refining the language’s capabilities. Achieving these research goals would substantially elevate \Qutes as an advanced and accessible quantum programming language, bridging the gap between high-level quantum algorithm design and low-level circuit implementation, with the potential to drive significant advancements in quantum computing.

\bibliographystyle{plain}
\bibliography{bibliography}

\begin{thebibliography}{10}

\bibitem{silq}
Benjamin Bichsel, Daniel Zhan, Goran Sutter, and Martin Vechev.
\newblock Silq: A high-level quantum language with safe uncomputation and
  intuitive semantics.
\newblock In {\em Proceedings of the 41st ACM SIGPLAN Conference on Programming
  Language Design and Implementation (PLDI)}, pages 286--300. ACM, 2020.

\bibitem{pyquil}
Rigetti Computing.
\newblock Pyquil: A python library for quantum programming.
\newblock \url{https://github.com/rigetti/pyquil}, 2023.
\newblock Accessed: 2024-11-19.

\bibitem{openqasm}
Andrew~W. Cross, Lev~S. Bishop, John~A. Smolin, and Jay~M. Gambetta.
\newblock A quantum assembly language.
\newblock {\em arXiv preprint arXiv:1707.03429}, 2017.

\bibitem{Deutsch1992RapidSO}
David Deutsch and Richard Jozsa.
\newblock Rapid solution of problems by quantum computation.
\newblock {\em Proceedings of the Royal Society of London. Series A:
  Mathematical and Physical Sciences}, 439:553 -- 558, 1992.

\bibitem{FPV24}
Simone Faro, Arianna Pavone, and Caterina Viola.
\newblock Families of constant-depth quantum circuits for rotations and
  permutations.
\newblock In {\em Proceedings of the 25nd Italian Conference on Theoretical
  Computer Science, Torino, Italy, September 11-13, 2024}, 2024.

\bibitem{Software}
Mark Fingerhuth.
\newblock Open-source quantum software projects.
\newblock 2024.

\bibitem{HSKRH24}
Hermann Fürntratt, Paul Schnabl, Florian Krebs, Roland Unterberger, and Herwig
  Zeiner.
\newblock {\em Towards Higher Abstraction Levels in Quantum Computing}, pages
  162--173.
\newblock 03 2024.

\bibitem{quipper}
Alexander~S. Green, Peter~LeFanu Lumsdaine, Neil~J. Ross, Peter Selinger, and
  Beno{\^\i}t Valiron.
\newblock Quipper: A scalable quantum programming language.
\newblock In {\em Proceedings of the 34th ACM SIGPLAN Conference on Programming
  Language Design and Implementation}, pages 333--342, New York, NY, USA, 2013.
  ACM.

\bibitem{qiskit}
Ali Javadi-Abhari, Matthew Treinish, Kevin Krsulich, Christopher~J. Wood, Jake
  Lishman, Julien Gacon, Simon Martiel, Paul~D. Nation, Lev~S. Bishop,
  Andrew~W. Cross, Blake~R. Johnson, and Jay~M. Gambetta.
\newblock Quantum computing with {Q}iskit, 2024.

\bibitem{qsharp}
Microsoft Quantum~Development Kit.
\newblock The q\# programming language.
\newblock \url{https://learn.microsoft.com/en-us/azure/quantum/}, 2023.
\newblock Accessed: 2024-11-19.

\bibitem{antlr}
Terence Parr.
\newblock {\em The Definitive ANTLR 4 Reference}.
\newblock Pragmatic Bookshelf, 2 edition, 2013.

\bibitem{Simulators}
Quantiki.
\newblock List of qc simulators.
\newblock 2024.

\bibitem{cirq}
Google~AI Quantum and Collaborators.
\newblock Cirq: A python framework for creating, editing, and invoking noisy
  intermediate scale quantum (nisq) circuits.
\newblock \url{https://quantumai.google/cirq}, 2023.
\newblock Accessed: 2024-11-19.

\bibitem{twist}
Charles Yuan, Christopher McNally, and Michael Carbin.
\newblock Twist: sound reasoning for purity and entanglement in quantum
  programs.
\newblock {\em Proc. ACM Program. Lang.}, 6(POPL), January 2022.

\bibitem{Zangi_2023}
Sultan~M. Zangi, Chitra Shukla, Atta ur~Rahman, and Bo~Zheng.
\newblock Entanglement swapping and swapped entanglement.
\newblock {\em Entropy}, 25(3):415, February 2023.

\end{thebibliography}

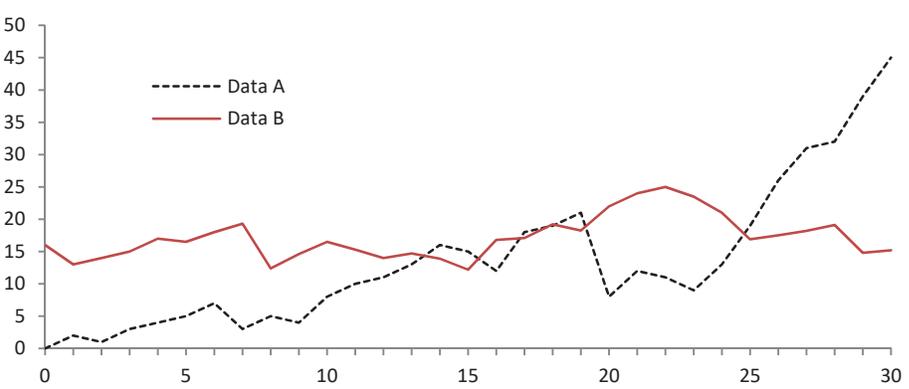
\end{document}